\documentstyle [12pt,osa,eqsecnum]{revtex}

\begin{document}
\thispagestyle{empty}
\title {
\bf Probing the Concept of Statistical Independence
\\of Intermediate-Mass Fragment Production\\ in Heavy-Ion Collisions}

\author{  W.\ Skulski,\cite{bySkul} J.\ T\~oke, and W.\ U.\ Schr\"oder}
\address{\em Dept.\ of Chemistry and NSRL, University of Rochester,
        Rochester, New York 14627 }

\maketitle

\begin{abstract}

It is studied to what extent the characteristics of multi-IMF
(intermediate-mass fragments) events can be derived from the
properties of the observed single-IMF transverse-energy  spectra. It
is found that the spectra of total transverse IMF energy ($E_t$) in
multi-IMF events are well represented by ``synthetic'' spectra
obtained by a multiple folding of the single-IMF transverse energy
spectrum. Further, it is shown that, using the experimental IMF
multiplicity distribution in the folding procedure, it is possible to
reproduce the observed trends in the IMF multiplicity distributions
for fixed values of the total transverse energy $E_t$. Accordingly,
the ``synthetic'' multiplicity distributions show a binomial
reducibility and an Arrhenius-like scaling similar to that reported in
the recent literature.  Consistent with the concept of statistical
independence of multiple IMF production, similar results are obtained
when the above folding-type synthesis is replaced with one based on
mixing of events with different IMF multiplicities. For statistically
independent IMF emission, the observed binomial reducibility and
Arrhenius-like scaling are shown to be merely reflections of the shape
of the single-IMF transverse-energy spectrum. Hence, a valid
interpretation of these IMF distributions in terms of a particular
production scenario has to explain independently the observed shape of
the single-IMF $E_t$ spectrum.

\end{abstract}
\bigskip
(PACS 25.70Lm, 24.10.Pa, 25.70.Mn, 25.70.Pq)

\newpage
\bigskip

\section{Introduction}

Recently, in several studies\cite{BerkPapers,Toke}  of
heavy-ion-induced nuclear multifragmentation it was concluded  that
certain characteristics of the intermediate-mass fragment (IMF) data
are suggestive of statistical independence of multiple IMF
production.\cite{Toke} Concepts of a  binomial or a Poissonian
reducibility\cite{BerkPapers} have been suggested to express the
particular observed independence in mathematical terms. The concept of
statistical independence or reducibility presumes that certain
characteristics of multi-IMF events are reflections of the single-IMF
production process. For example, the concept of a binomial
reducibility\cite{BerkPapers} presumes that the probability of
multiple IMF production is a reflection of the single-IMF production
probability - the two quantities of interest being connected via the
expression for a binomial distribution:

\begin{equation}
\qquad P_n^m(p) = {m!\over n!(m-n)!}p^n(1-p)^{m-n}\;,
\label{eq_bindist}
\end{equation}

\noindent
where $P_n^m(p)$ is the multiple-IMF production probability and
the parameters $m$ and $p$ are the number of (binomial) tries and
the probability for success in any of these trials, respectively.

Although so far, no theoretical model has been proposed that would
explain a binomial reducibility, one would reasonably expect  that a
successful model would at the same time link other characteristics of
multi-IMF events to the characteristics of single-IMF production. With
these expectations in mind, the present paper investigates, to what
extent the empirical systematic of multiple IMF production can be
derived from empirical characteristics of single-IMF production and,
specifically, from the single-IMF transverse-energy spectrum. It is
assumed in this study that the hypothesis of an independent IMF
production holds and that, accordingly, the total transverse-energy
spectra of multiple IMFs can be synthesized from the transverse-energy
spectrum of a single IMF.  Section 2 presents such a synthetic
procedure, based on multiple folding of a schematic single-IMF
spectrum. Section 3 discusses results obtained when empirical
single-IMF spectra are substituted for the schematic distributions
used in Section 2, and Section 4 discusses effects of an inclusion of
transverse energies of light charged particles in the definition of
the total transverse energy. An alternative synthesis method based on
fragment swapping across different events of different IMF
multiplicity is discussed in Section 5. Section 6 summarizes the
findings of the present study.

\section{Synthesis of Multi-Fragment $E_{t}$ distributions}

In the following, the total transverse energy of charged reaction
products is defined as \begin{equation} \qquad
E_t=\Sigma_{i=1}^{mult}E_isin^2(\Theta_i), \label{Et} \end{equation}
where the summation extends over all particles with atomic numbers
$1\leq Z\leq 20$. In the present section, only IMFs with  $3\leq Z\leq
20$ are included in the definition of $E_t$. Effects of an inclusion
of light charged particles are discussed further below, in Section
4.The following empirical characteristics of multi-IMF events
represent the rationale behind the synthesis procedure presented in
this section: 

\begin{enumerate} 

\item{The $E_{t}$ distributions of individual IMFs are approximately
exponential, with inverse logarithmic-slope parameters $T$ of the
order of 30 to 50 MeV.\cite{Toke,TokeSnowbird,Wieloch,Wielocha} As an
example, the experimental transverse-energy spectrum is shown in
Fig.~1, for single IMFs from the Bi+Xe reaction\cite{Toke} at $E/A=28$
MeV. Since $E_{t}$ is an angle-integrated quantity, the corresponding
spectrum may differ from a Maxwell-Boltzmann shape. Superimposed on
the data are an exponential distribution with $T=50$ MeV (dotted line)
and a two--parameter Gaussian fit (solid line) to the spectrum. While
the exponential function deviates from the experimental spectrum both
at high and low values of $E_t$, it simplifies the following
demonstration of qualitative features of total IMF $E_t$ spectra. }

\vspace*{3.7in}
\includegraphics{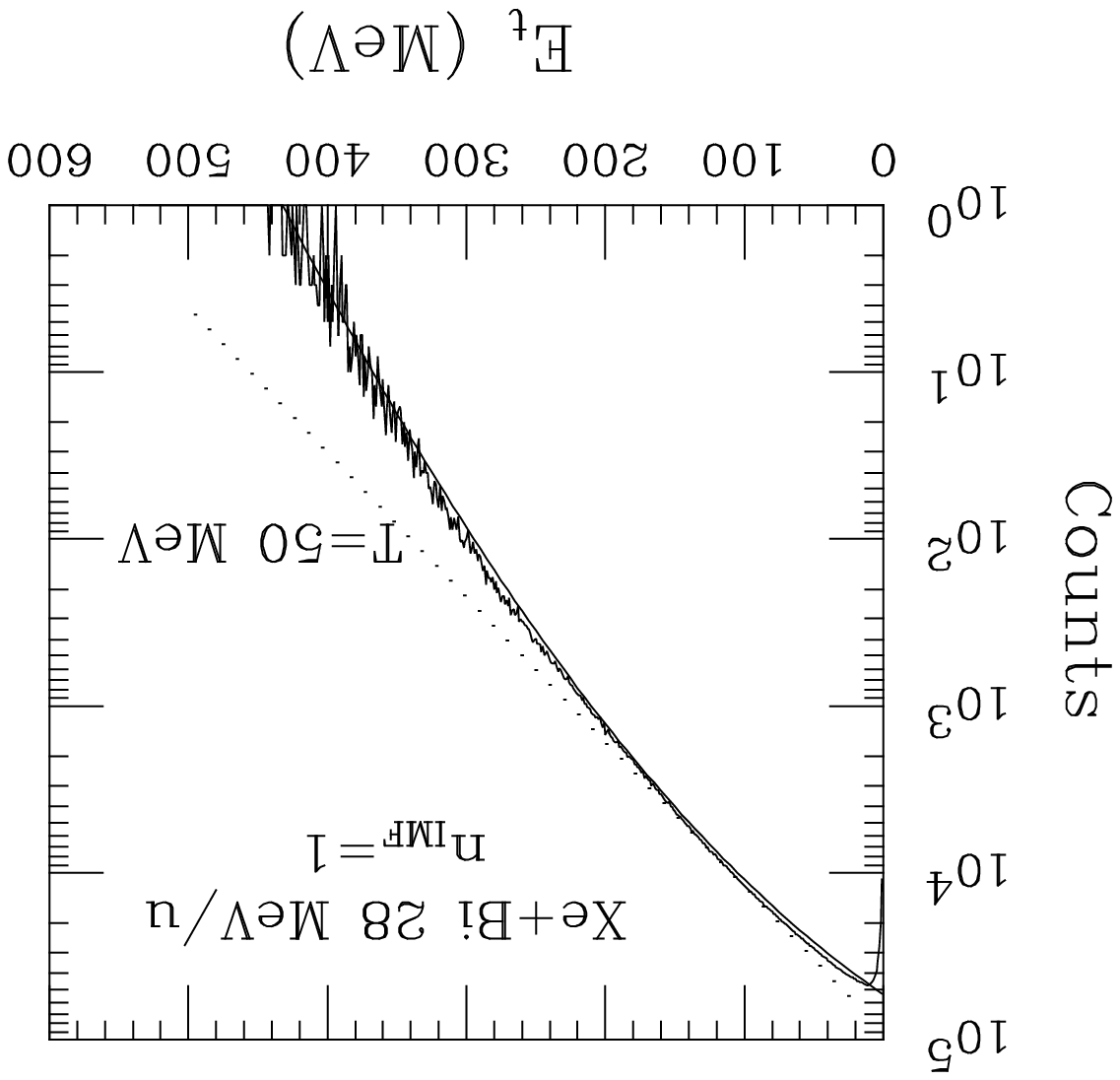}

{\footnotesize
\begin{quotation}
\noindent

Fig. 1. Intermediate-mass fragment transverse-energy spectrum of
single-IMF events. The dotted line is an exponential distribution,
$\exp{(-E_t/T)}$, used as a first--order approximation of single--IMF
spectra in Eq.~\ref{eq:expo}. The spectrum itself was used as a model
distribution $\omega(e)$ for a numerical evaluation of the
integrals~(\ref{eq:FoldingIMFs}) and (\ref{eq:FoldingLCPs}) with a
Monte Carlo method. \end{quotation}} 

\item{To a good approximation, $E_{t}$ spectra of individual IMFs are
independent of the IMF coincidence fold (multiplicity)
$n$.\cite{Toke,TokeSnowbird,Wieloch,Wielocha} Although discernible, the
dependence on $n$ is weak.} 

\item{In addition to IMFs, one also observes light charged particles
(LCPs), which greatly outnumber IMFs, for every $n$. Even for $n=0$,
the average multiplicity of LCPs is of the order of 10 and, hence,
quite significant. } 

\item{At low bombarding energies, e.g., for $E/A\approx 30$ MeV, the
IMF contribution can dominate the total transverse energy $E_t$, for
all but the lowest multiplicities
$n$.\cite{Toke,WSAnnReport96,PhairPriv} For larger bombarding
energies, the relative IMF contribution decreases but remains always
significant.} 

\end{enumerate}

Assuming a statistically independent IMF production, the probability
distribution $\Omega_n(e_1, ..., e_n)$ for events with IMF
multiplicity $n$ can be expressed in the form of a product of $n$
individual single--IMF distributions $\omega_i(e_i)$

\begin{equation} 
\label{eq:Independence} 
\Omega_n (e_1, ..., e_n) =
\prod_{i=1}^{n} \omega_i (e_i)  \approx \prod_{i=1}^{n} \omega (e_i).
\end{equation}

In Eq.~\ref{eq:Independence}, as in the following discussion,
single--particle variables will be denoted by lower-case characters,
while upper-case characters indicate multi--particle variables or
distributions. For example, the symbol $\Omega$ will be used to denote
multi--particle probability distributions, while the symbol $\omega$
indicates the corresponding single--particle distribution. Similarly,
the variable $e_i$ denotes the contribution of particle number $i$ to
the total (multi-particle) transverse energy $E_{t}$. While the first
part of Eq.~\ref{eq:Independence} represents the statistical
independence, the second expresses the equality of all individual
energy distributions.

Given the above form (\ref{eq:Independence}) of $\Omega_n(e_1, ...,
e_n)$, the distribution  $\tilde{\Omega}_n(E_t)=\tilde{\Omega}_n
(e_1+\cdots +e_n)$ of the total transverse energy can be expressed as

\begin{equation} 
\label{eq:GeneralFolding} 
\tilde{\Omega}_n (E_t)    = \int
dt_1\cdots \int dt_{n-1} \Omega_n(t_1, ..., t_n) \cdot \delta (E_t - t_1 -
\cdots - t_n), 
\end{equation}

where the integration is carried out over a hyperplane of a
constant $E_{t}$, as enforced by the $\delta$ function.
Equation~\ref{eq:GeneralFolding} can be further simplified to:

\begin{equation}
\label{eq:FoldingIMFs}
\tilde{\Omega}_n (E_t) =
\int_{0}^{E_t} \omega (E_t-t_1)dt_1 \int_{0}^{t_1}  \omega (t_1-t_2)dt_2
\cdots
\int_{0}^{t_{n-2}}  \omega(t_{n-2}-t_{n-1}) \omega (t_{n-1}) dt_{n-1}
\end{equation}

In the general case of an arbitrary individual single-particle
distribution $\omega$,  the integration in Eq.~\ref{eq:FoldingIMFs}
can be performed numerically, at least for reasonably low IMF
multiplicities.  When a simple exponential function is taken to
approximate $\omega$,  this integral can be evaluated analytically for
arbitrarily large IMF multiplicities:

\begin{equation}
\label{eq:expo}
\omega (e_i) = \beta \cdot \exp{ (-\beta \cdot e_i)}
\end{equation}

\begin{equation}
\tilde{\Omega}_n (E_t) =  \beta \cdot \frac{1}{(n-1)!} \cdot
(\beta E_t)^{n-1}  \cdot \exp{ ( -\beta \cdot E_t) }  \label{eq:Poiss}
\end{equation}

In principle, the inverse slope parameter $\beta=1/T$ could depend on
the IMF multiplicity $n$, but the experimental dependence on $n$ is
known to be weak.\cite{TokeSnowbird,Wieloch,Wielocha}  According to
Eq.~\ref{eq:Poiss}, to the extent that the single-IMF
transverse-energy spectra can be approximated by exponentials, the
distribution of the composite observable $E_{t}$ is of Poissonian form
for each IMF coincidence order $n$.

In order to calculate quantitatively the two-dimensional joint
probability distribution $P(n,E_t)$ for a given reaction, one should
combine the one-dimensional $E_{t}$ distributions (\ref{eq:Poiss}) for
given $n$-values with the IMF multiplicity distribution $y_n$,
measured for this reaction, i.e.,

\begin{equation}
   \label{eq:WeightedPoiss}
   P(n,E_t) =  y_n \cdot \tilde{\Omega}_n (E_t)
\end{equation}

Typical experimental $y_n$ distributions are depicted in  Fig.~2. For
the experimental data discussed here, they can be approximated well by
Gaussians, $y_n\propto \exp{(-(n-1)^2/2\cdot \sigma^2)}$  (solid line
in the figure).

\vspace*{3.7in}
\includegraphics{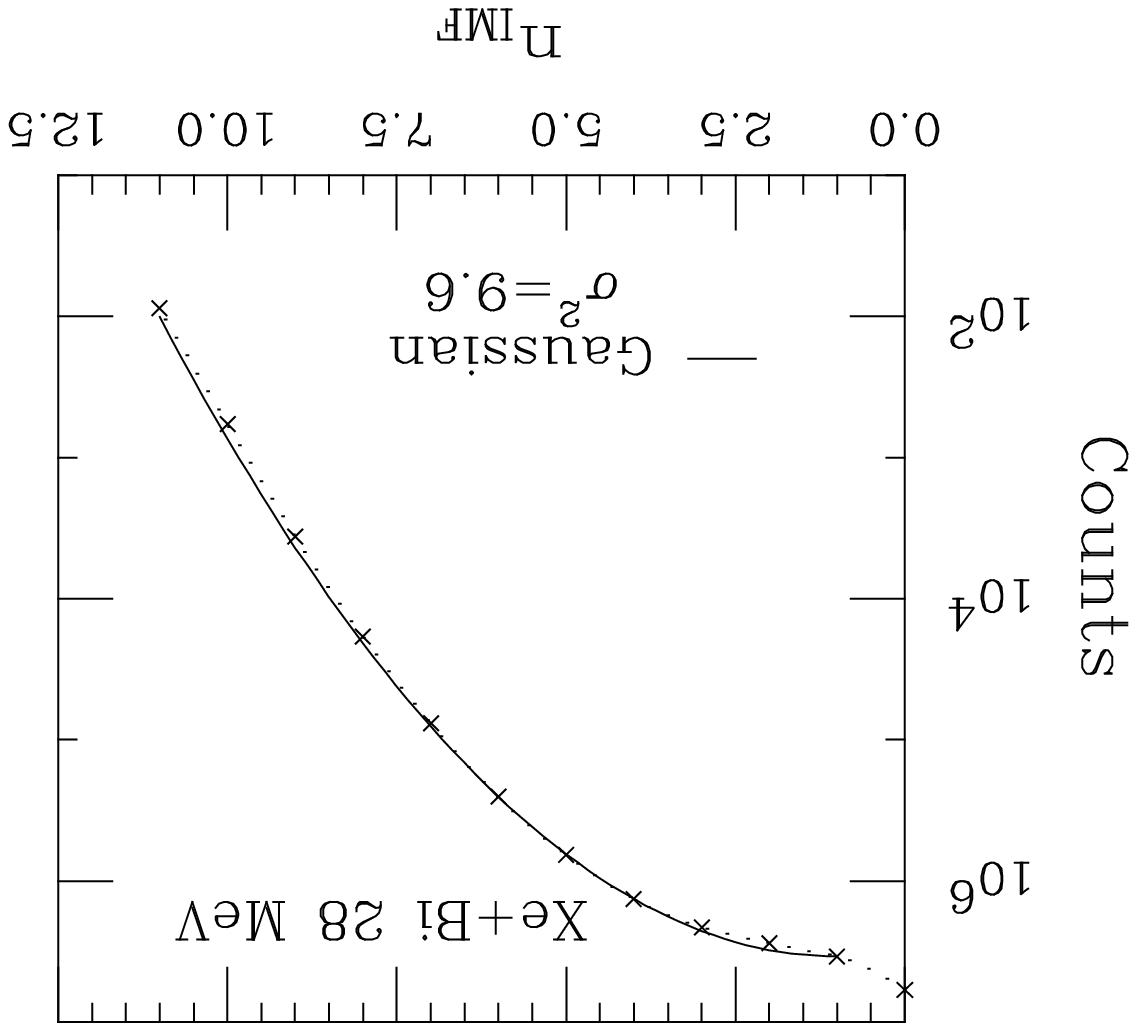}

{\footnotesize 
\begin{quotation}
\noindent 
Fig.~2. Intermediate-mass fragment yield distribution of $n$--fold IMF
events, for $n=0,1,2,\ldots$. The solid line is a Gaussian
distribution, $y_n\propto \exp[-(n-1)^2/2\cdot \sigma^2]$. This yield
distribution was used for a numerical evaluation of the
integrals~(\ref{eq:FoldingIMFs}) and (\ref{eq:FoldingLCPs}) with a
Monte Carlo method.  \end{quotation}}

\bigskip
The joint distribution in Eq.~\ref{eq:WeightedPoiss}
allows one to synthesize conditional IMF multiplicity distributions,
gated with different values of $E_t$, and to calculate the binomial
parameters $m$ and $p$ for these distributions for further analysis
along the lines adopted in a series of recent papers.\cite{BerkPapers}
Specifically, one calculates

\begin{equation}
   \label{eq:meann}
   \overline{n} =
   \sum_{i=1}^{N} i \cdot P(i,E_t) /
   \sum_{i=1}^{N} P(i,E_t) ,
\end{equation}

\begin{equation}
\label{eq:sigman}
\sigma_n^2 =  \sum_{i=1}^{N}
(i-\overline{n})^2 \cdot P(i,E_t)/  \sum_{i=1}^{N}
\ P(i,E_t),
\end{equation}

\begin{equation}
m = \overline{n} / p ,  \label{eq:m}
\end{equation}

\begin{equation}
p = 1 - \sigma_n^2 / \overline{n}.   \label{eq:p}
\end{equation}

\vspace*{5.2in}
\includegraphics{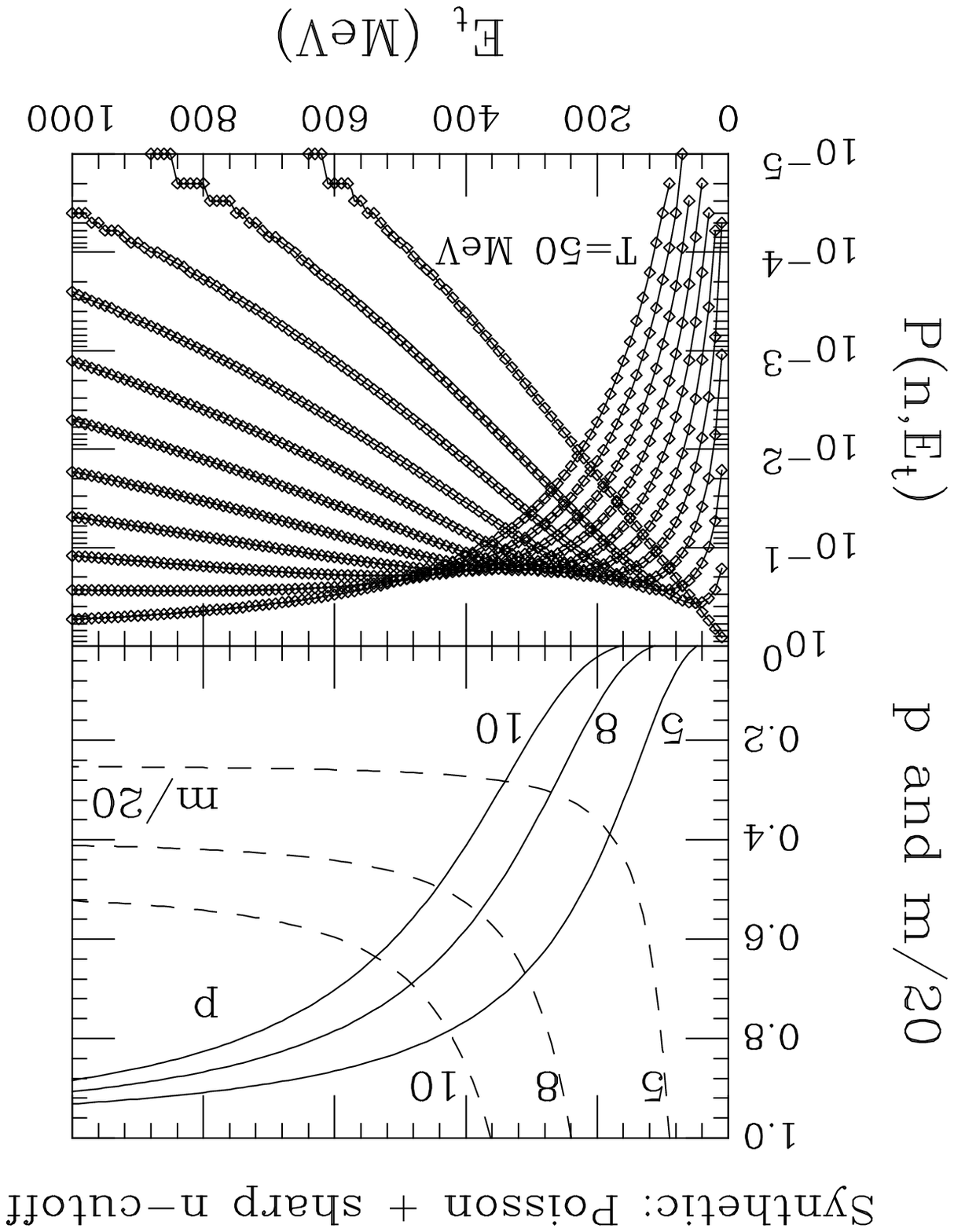}
{\footnotesize 
\begin{quotation} 
\noindent
Fig. 3. Lower panel: The distributions~(\ref{eq:Poiss}) are plotted up
to the order of $N=11$. Note the qualitative similarity of the
distributions~(\ref{eq:Poiss}) and the $E_t$ distributions due to
multifragmentation, plotted in the subsequent figures.  Upper panel:
The values of parameters $p$ and $m$ calculated with
equations~(\ref{eq:p}) and (\ref{eq:m}) and
distributions~(\ref{eq:Poiss}), with sharp--cutoff weights $y_n=1$,
for n up to $N$, and $y_n=0$ for $n>N$. Three example values of
$N=5,8,10$ are indicated in the figure. Note the severity of the
truncation effect, which causes the calculated $p$ and $m$ to deviate
from their limiting values, zero and infinity, respectively. The
limits are reached only for $N$ approaching infinity. 
\end{quotation}}

\bigskip
It is the use of the constrained empirical distribution $y_n$ that
leads to a reduction of the variances of the conditional multiplicity
distributions below the Poissonian limit of $\overline{n} =\sigma_n^2$
and, hence, ensures positive values for the parameters $m$ and $p$.
This is illustrated in Fig.~3, where a schematic sharp--cutoff
distribution of $y_n$ serves to demonstrate the importance of
truncation. Note the large magnitude of the truncation effect, which
causes the calculated parameters $p$ and $m$ to deviate from their
limiting values, zero and infinity, respectively. These limits are
reached only when the number of folds approaches infinity. Significant
values for both $p$ and $m$ have been generated within the physically
relevant range of $E_t$ by restricting the number of folds to $\approx
10$, i.e., a number of coincidence folds typically measured in an
experiment.

\vspace*{5.2in}
\includegraphics{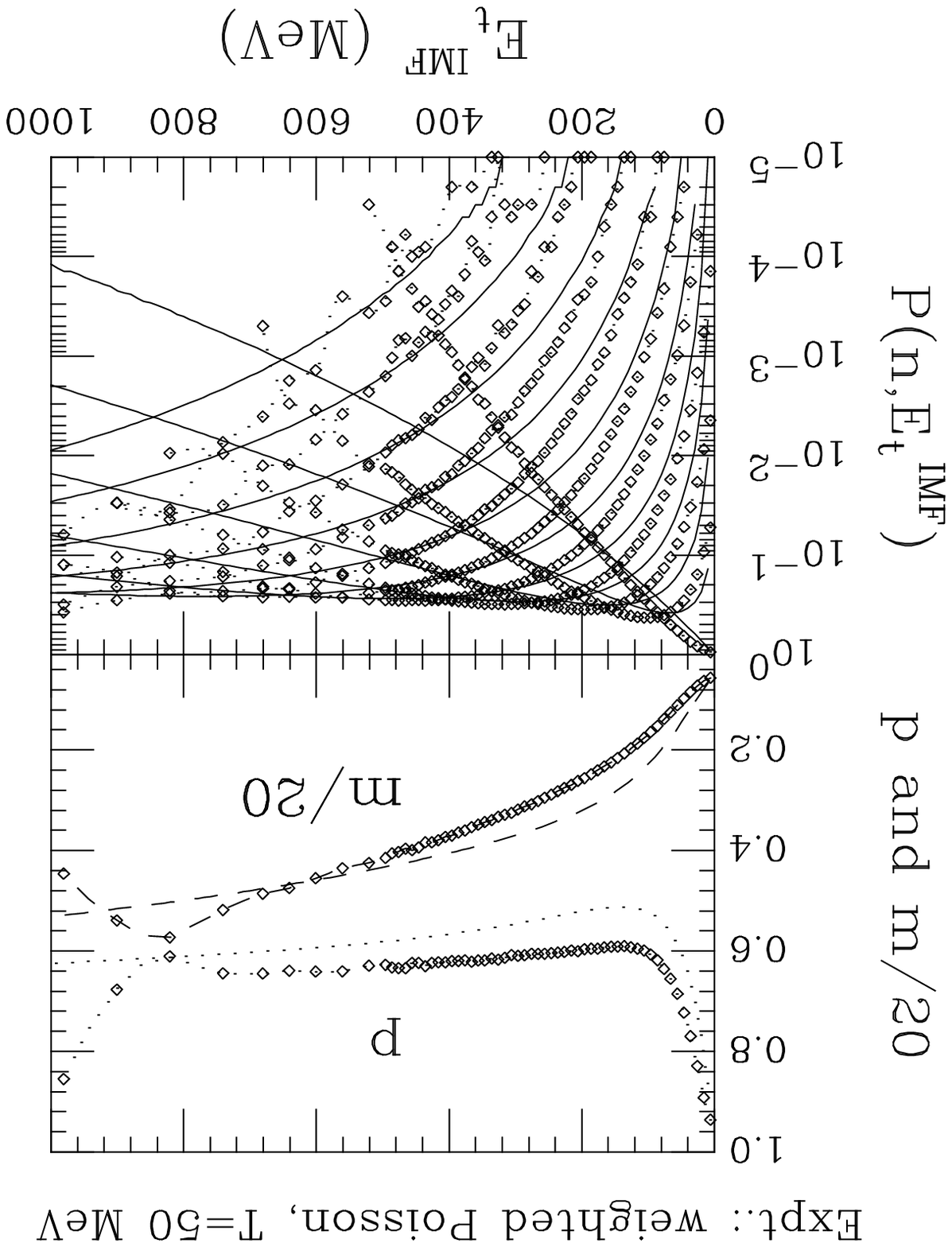}
{\footnotesize 
\begin{quotation}
\noindent
Fig.~4. Lower panel: The $E_t$ distributions~(\ref{eq:WeightedPoiss})
weighted with the experimental IMF yields $y_n$ are compared with
experimental IMF--only transverse-energy distributions. The upper part
of the figure shows the respective $p$ and $m$ parameters as
calculated with Eqs.~(\ref{eq:p}) and (\ref{eq:m}), respectively.
\end{quotation}}

\bigskip
The synthetic distributions are compared in Fig.~4 with experimental
IMF transverse-energy distributions. In the calculations, the IMF
yields $y_n$ were taken from experiment.\cite{TokeArrhenius} The
inverse slope parameter $\beta^{-1}=50$ MeV of the IMF spectra was
adjusted to approximately match the singles ($n=1$) spectrum up to
about 200 MeV (see Fig.~1). As seen in this figure, already a
synthesis using an exponential singles spectrum reproduces
qualitatively the gross features of the experimental $E_{t}$
distributions. The lack of a quantitative agreement with experimental
data is not surprising, since the used schematic exponential
single--particle distribution~(\ref{eq:expo}) is only a very crude
approximation of the actual single--particle spectrum. As seen in
Fig.~1, the tail of the exponential function does not match the $n=1$
spectrum for $E_t>200$ MeV. Nevertheles s, the overall features of the
experimental $E_{t}$ distributions are reproduced qualitatively by
those calculated with Eq.~\ref{eq:WeightedPoiss}.

The upper part of Fig.~4 presents a comparison between the parameters
$p$ and $m$, extracted from experimental data, with the ones
calculated from Eqs.~\ref{eq:meann} and \ref{eq:sigman}, using the
distributions~(\ref{eq:WeightedPoiss}) displayed in Fig.~2. The
predicted values of $p$ and $m$ match the experiment remarkably well,
even though the calculated $E_{t}$ distributions in the lower panel
deviate from the experimental ones. This points towards the need for a
better description of experimental single--fragment $E_{t}$
distributions $\omega(e)$. More realistic shapes of these
single--particle distributions will be considered next.

\section{Synthesis with Realistic Single--IMF Distributions}

A more realistic rendering of the shape of the single--IMF
distribution than achieved by an exponential is provided by a
two--parameter Gaussian fit of the kind depicted by the solid line in
Fig.~1. Such a Gaussian spectrum was used in a numerical evaluation of
the integrals~(\ref{eq:FoldingIMFs}), employing a Monte-Carlo sampling
method results of which  are displayed in Fig.~5.  As seen in Fig.~5,
a remarkable quantitative agreement with experimental data is
achieved. This time, also the parameters $p$ and $m$ have been
reproduced. It is worth emphasizing, that only four parameters, two
shape parameters of the single-particle spectrum (cf. Fig.~1) and two
parameters describing the IMF yield distributions (cf. Fig.~2)) are
sufficient to reproduce the shapes of the multi--IMF $E_t$
distributions, and the $E_t$-dependencies of the parameters $p$ and
$m$.

\section{The Role of LCP Contributions}

In order to better represent the experimental situation, the
contribution by light charged particles to the total kinetic energy
has to be considered. This can be achieved easily by extending the
concept of a statistically independent fragment production to the LCP
emission. In this case,

\begin{equation}   \label{eq:FoldingLCPs}
   \tilde{\Omega}_n (E_t) =
   \tilde{\Omega} (e_0+e_1+,...,+e_n ) =
   \int_{0}^{E_t} \omega_0 (E_t-t_0)dt_0
   \int_{0}^{t_0}  \omega (t_0-t_1)dt_1 ....
\end{equation}

The index 0 now denotes the LCP contribution to the total $E_{t}$.
Since there are many LCPs contributing independently (even for the
case of no IMFs, $n=0$), according to the Central Limit Theorem, the
contribution of all LCP's, taken together, is approximately a 
Gaussian

\begin{equation}
   \label{eq:gauss}
   \omega_0 (e_0)  \propto
   \exp{ [ -(e_0 - \overline{e_0})^2 / 2\sigma_0^2 ]},
\end{equation}

\noindent
where the parameters $\overline{e_0}$ and $\sigma_0$ may depend on the
IMF multiplicity $n$. 

\vspace*{5.3in}
\includegraphics{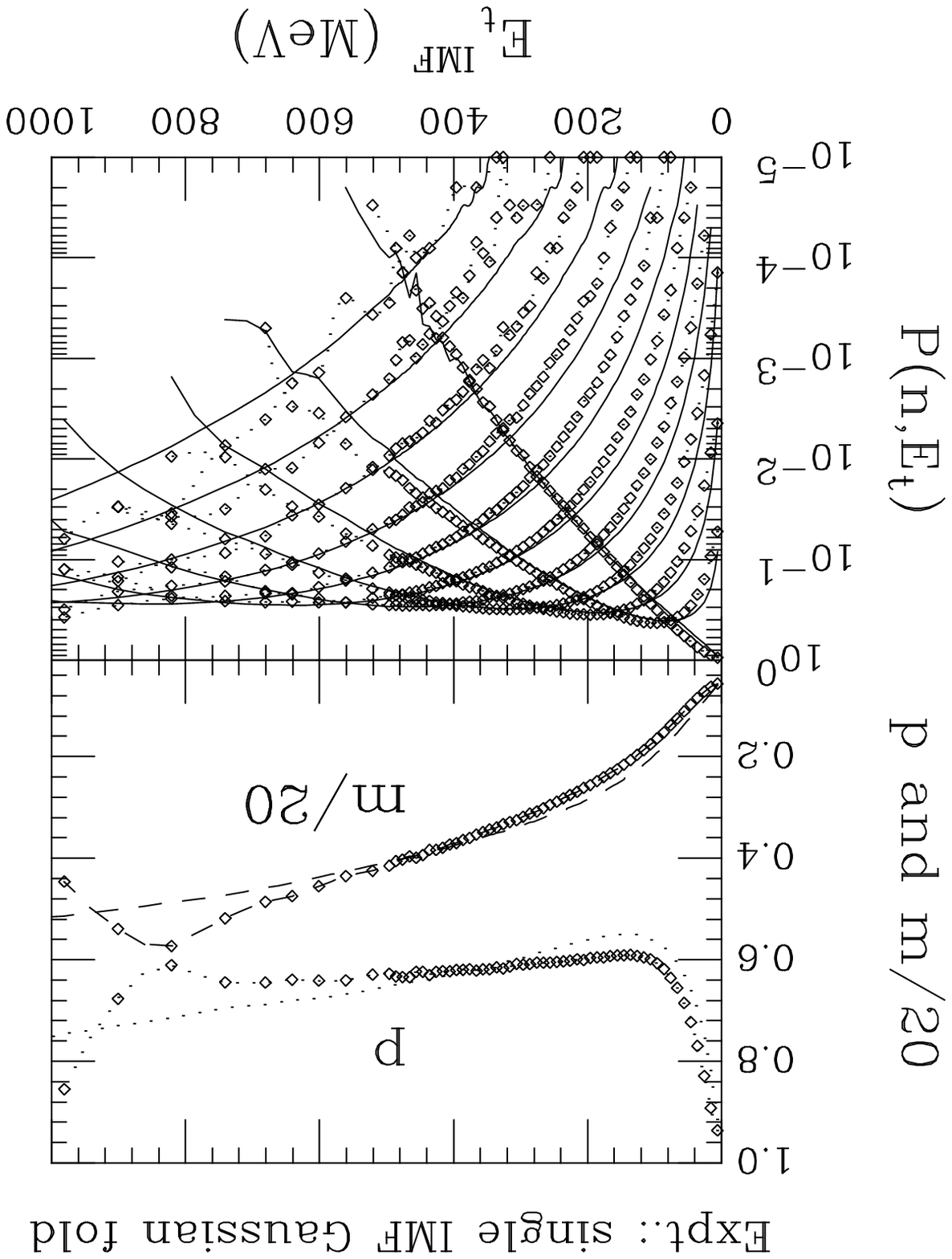}
{\footnotesize 
\begin{quotation}
\noindent 
Fig.~5. Lower panel: IMF transverse- energy distributions (solid
lines), predicted by folding a Gaussian form of the $n=1$ IMF
distribution multiple times with itself (without the LCP
contribution), according to formula~(\ref{eq:FoldingIMFs}). The
resulting $E_t$ spectra were weighted with the Gaussian IMF yields
$y_n$ from Fig.~2. The symbols connected with dotted lines represent
experimental data. Upper panel: $p$ and $m$ parameters extracted from
the experimental data (symbols connected with lines), compared with
the ones extracted from the set of folded $E_t$ distributions.
\end{quotation}}
\bigskip

Wieloch et. al.\cite{Wieloch,Wielocha}  have
considered the particular parameterization of $\overline{e_0}\propto
n$ and $\sigma_0 = const$. The present paper employs values of
$\overline{e_0}$ and $\sigma_0$ taken from
experiment,\cite{TokeArrhenius} namely $\overline{e_0}\propto n$ for
$n=0,1,2$ and $\overline{e_0}\approx const$ for $n\geq 2$. With this
choice of parameter values $\overline{e_0}$ and $\sigma_0$, the
integral~(\ref{eq:FoldingLCPs}) was evaluated numerically in the same
way as before. The results are displayed in Fig.~6, showing
satisfactory agreement with experimental data. The parameters $p$ and
$m$ are also rather well reproduced by the calculation, except for the
deviations from experimental data at $E_t<100$ MeV. The deviations are
attributed to an imperfect representation of LCP spectra by Gaussians,
and in particular to the intense contributions of peripheral events to
the LCP spectra at low $E_t$. It should be noted, that low--$E_t$ data
are usually considered less important in the context of
multifragmentation.

\vspace*{5.3in}
\includegraphics{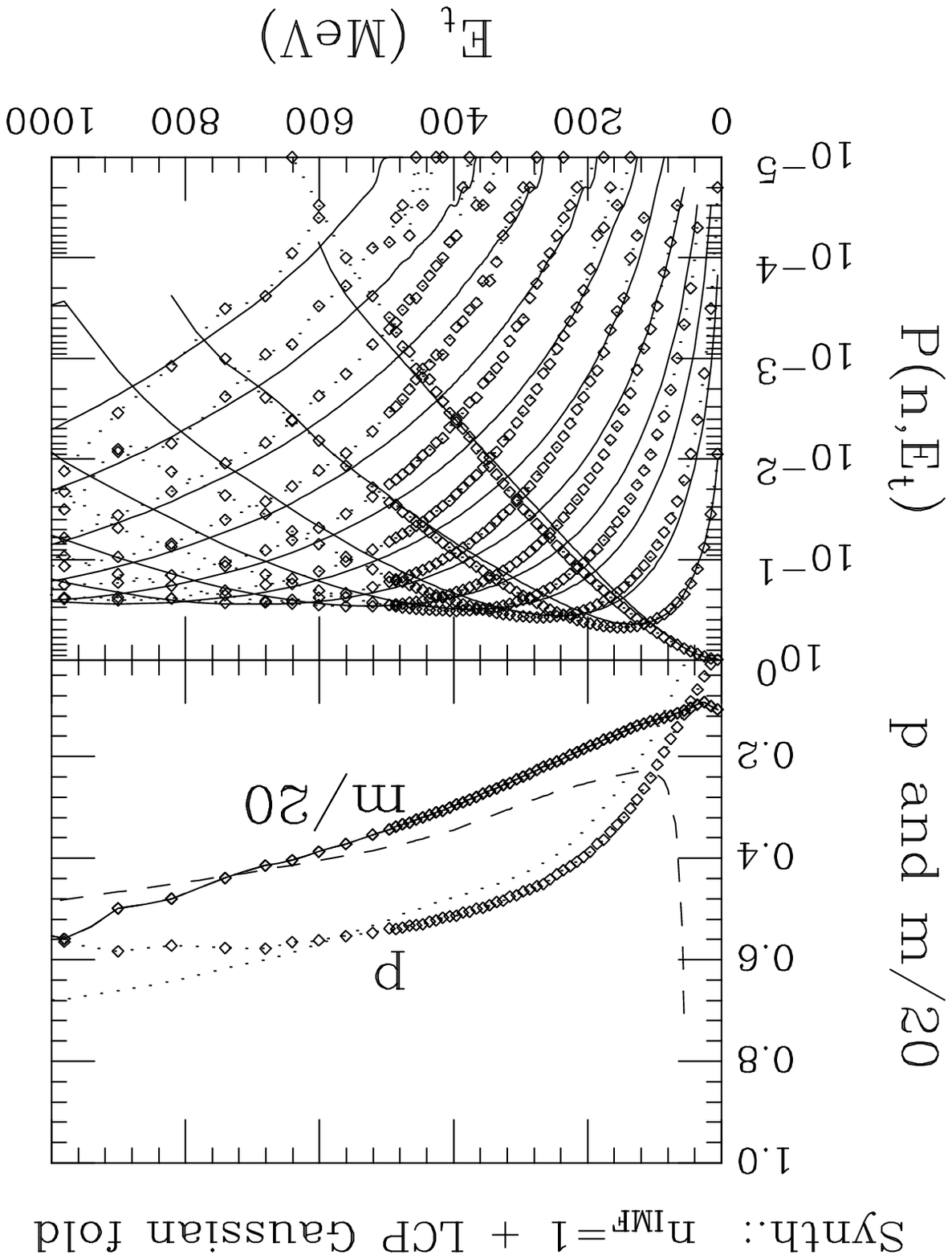}
{\footnotesize 
\begin{quotation} 
\noindent
Fig.~6. Lower panel: IMF transverse-
energy distributions (solid lines), predicted by folding a Gaussian
approximation to the $E_t$ distribution for $n=1$ multiple times with itself
and with the LCP contributions, according to formula~(\ref{eq:FoldingLCPs}).
The resulting $E_t$ spectra were weighted with the yields $y_n$ from
Fig.~3. The symbols connected with dotted lines represent
experimental data. Upper panel: Experimental $p$ and $m$ parameters are
compared with those predicted by folding. 
\end{quotation}}

\section{Mixed-event calculations}

As an alternative approach to synthesizing multi-IMF events, the
experimental data\cite{TokeArrhenius} have been subjected to ``event
mixing'', a procedure where both IMFs and LCPs are randomly exchanged
between different events. For every event, the number of LCPs,
$n_{LCP}$, and the number of IMFs, $n$, were preserved, but not the
identities of these particles. Both IMFs and LCPs were swapped among
events of all multiplicities (regarded as a common ``fragment pool''),
in order to ensure a complete ``loss of memory'' in the synthetic
mixed--event data set.

The above random particle exchange between events ensured that
multi-particle correlations were erased from experimental data, while
preserving approximately the shape of the single-particle
distributions. In this fashion, a statistical independence among
particles was strictly imposed on experimental data, even if some
degree of multi-particle correlations had been present in the original
data set. This procedure ensured that the assumptions underlying
Eq.~\ref{eq:FoldingLCPs} were strictly fulfilled by the data. It
should be stressed, that event mixing does not represent any
additional assumption brought into the problem. Rather, it is a
convenient way to evaluate the integral ~(\ref{eq:FoldingLCPs}), where
all single--particle probability distributions were sampled directly
from the experimental data, without recourse to a particular
parameterization.

\vspace*{5.2in}
\includegraphics{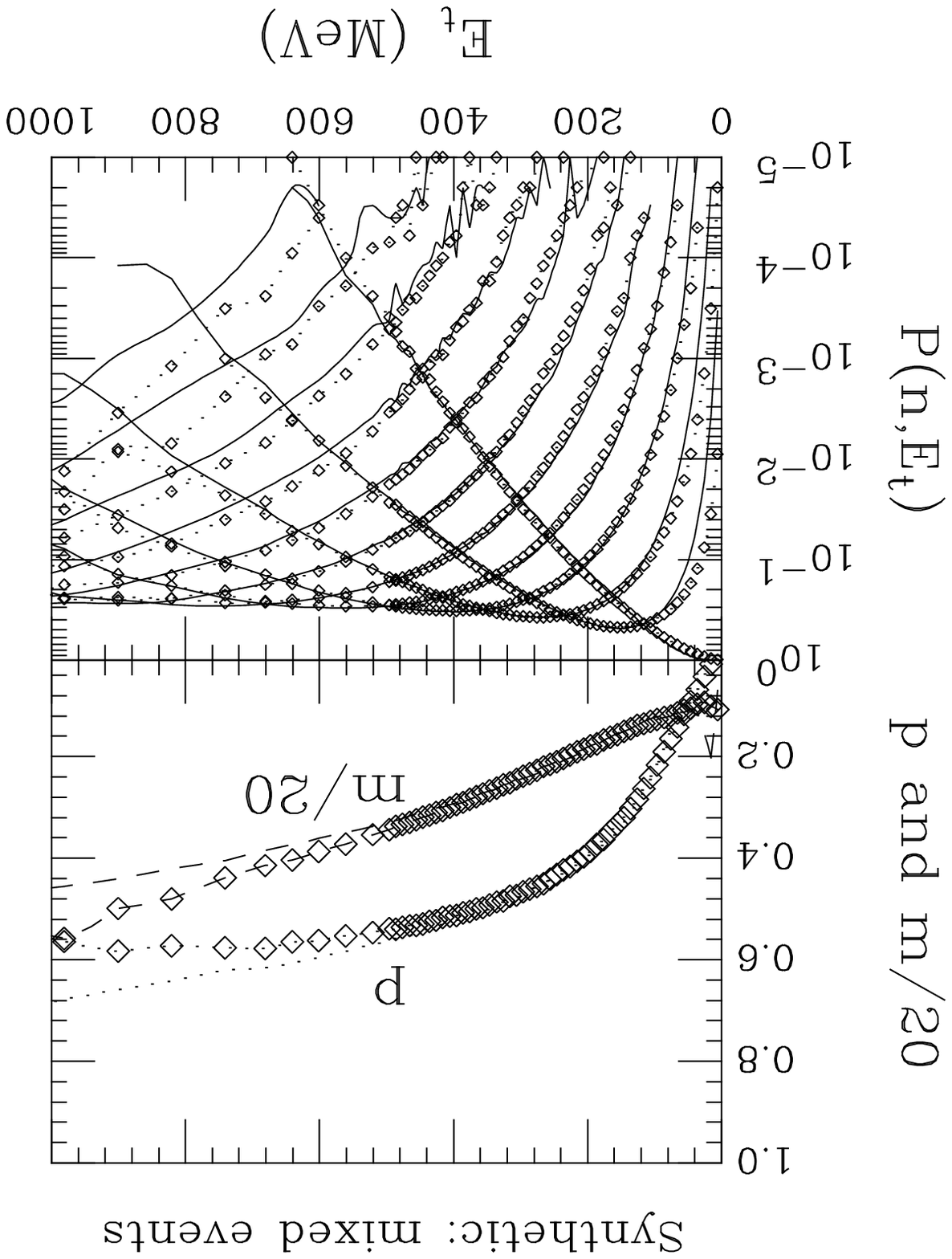}
{\footnotesize 
\begin{quotation} 
\noindent
Fig.~7. Transverse-energy distributions
(lower panel) and $p$ and $m$ parameters (symbols with dotted line in upper
panel) extracted from the experimental data set and from the same set, but
subjected to random particle swapping (solid lines). 
\end{quotation}}

\bigskip
Subsequently, the synthetic events were subjected to the same analysis
procedure as the original data. The results of this mixed-event
analysis are compared with the original data in Figs.~7 and 8. Very
good agreement between both sets of results is observed from these
figures,  both with regard to the $E_{t}$ distributions and the
extracted parameters $p$ and $m$. This result confirms again that the
concept of a statistically independent IMF production is a valid one.

\vspace*{3.9in}
\includegraphics{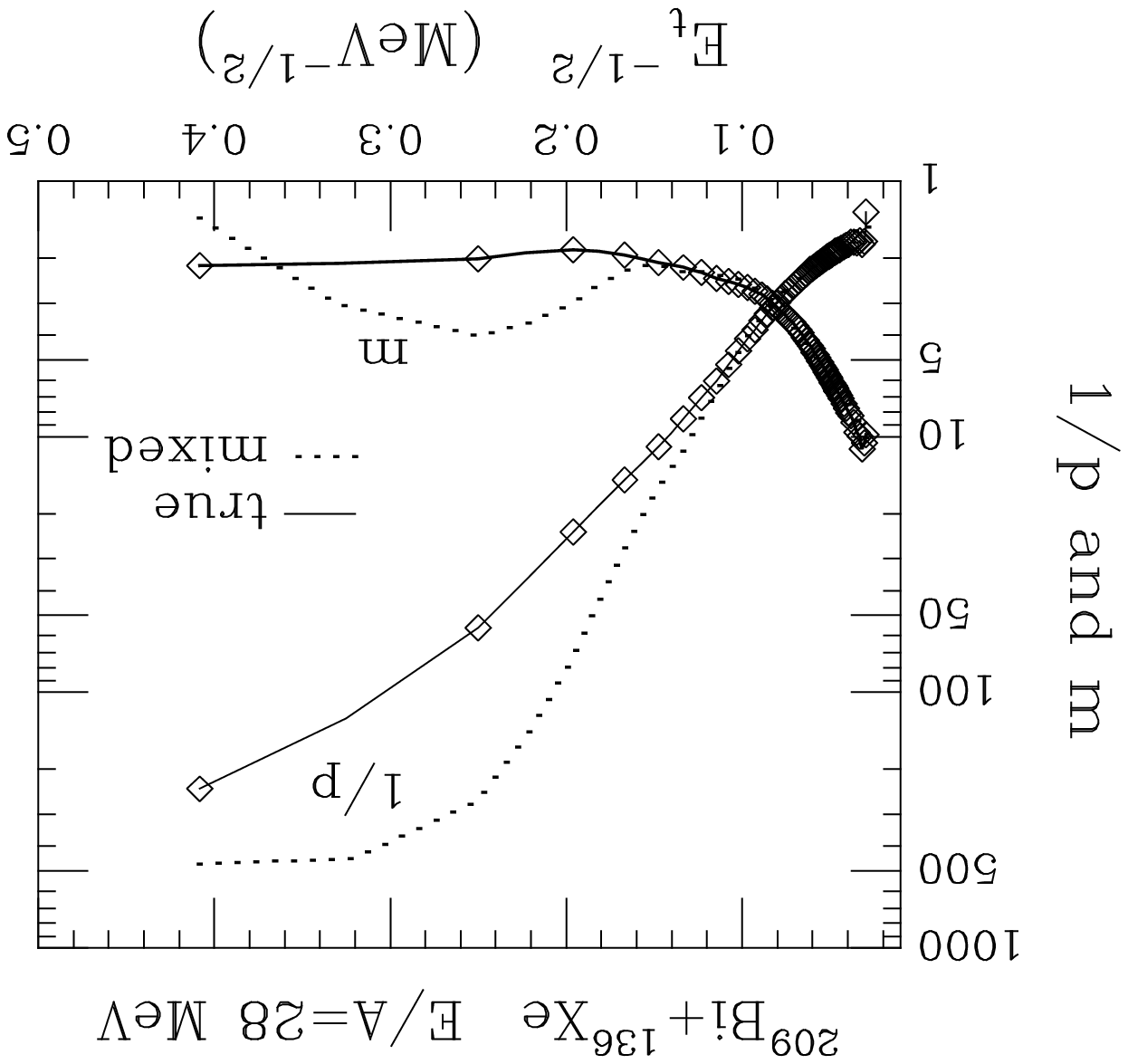}
{\footnotesize 
\begin{quotation} 
\noindent
Fig.~8. The $p$ and $m$ parameters
extracted from the experimental data set (symbols connected with solid line)
and from the same set subjected to random particle swapping (dotted lines),
plotted in an Arrhenius--like representation as functions of $E_t^{-1/2}$. 
\end{quotation}}

\bigskip

\section{Summary}

Accepting the validity of the findings of earlier
papers\cite{BerkPapers,Toke} regarding the statistical independence,
or reducibility, of multiple intermediate-mass fragment production,
one finds that important characteristics of multi-IMF events can be
derived already from the experimental shapes of single-IMF transverse
energy distributions. The characteristics of the synthetic events
generated from these distributions show remarkable resemblance to
those of the experimental multi-IMF events. The latter characteristics
include the binomial reducibility and an Arrhenius-like scaling
discovered in recent studies of nuclear
multifragmentation.\cite{BerkPapers}

Although the findings of the present study do not directly favor one
IMF production scenario over another, they do so indirectly. They draw
attention to the crucial importance of an understanding of the shapes
of the single-IMF $E_t$ distributions. In the present analysis, the
apparent binomial reducibility and the apparent Arrhenius-like scaling
are secondary phenomena, both reflecting the ``input'' shape of the
elementary $E_t$ distribution. Hence, a valid interpretation of the
significance of such reducibility and scaling in terms of a particular
IMF production scenario requires an independent explanation of the
underlying single-IMF $E_t$ distribution.

A purely thermal scenario, which could potentially justify
Arrhenius-like correlations between emission probability and
temperature of the system, makes quite definite predictions regarding
the elementary $E_t$ distributions, which should be very similar to
those of the lighter particles emitted from the same sources. These
predictions disagree with the experimental IMF spectra, which are
significantly harder than the light-particle $E_t$ spectra. This
observation justifies some skepticism towards an
interpretation\cite{BerkPapers} of Arrhenius-like plots in terms of
thermal scaling.

\section{Acknowledgements}

The Bi+Xe experiment at E/A = 28 MeV was conducted in collaboration with
S.P.\
Baldwin,   B.\ M.\ Quednau, L.G.\ Sobotka, J.\ Barreto, R.\ J.\ Charity, L.\
Gallamore, D.G.\ Sarantites, D.\ W.\ Stracener,  R.\ T.\ de Souza, and B.\
Lott.  Their contribution is highly appreciated.

 This work was supported by the U.\ S.\ Department\ of Energy Grant No.
DE--FG02--88ER40414.

\end{document}